\documentclass[osajnl,twocolumn,showpacs,superscriptaddress,10pt]{revtex4-1} %%
\usepackage{amsmath,amssymb,graphicx}
\usepackage{url}
\usepackage{graphicx}  
\usepackage{epstopdf}  
\usepackage{float}  
\usepackage[ruled,vlined]{algorithm2e} \DontPrintSemicolon
\usepackage{algpseudocode} \usepackage{multirow}  
\usepackage{array} 
\usepackage[export]{adjustbox}
\usepackage[ruled,vlined]{algorithm2e} \DontPrintSemicolon
\usepackage{algpseudocode} \usepackage{booktabs} \usepackage{tabu}
\makeatletter \newcommand{\removelatexerror}{\let\@latex@error\@gobble}
\makeatother

\usepackage{footnote}
\DeclareMathSymbol{\bot}{\mathord}{symbols}{"3F}
\DeclareMathSymbol{\perp}{\mathrel}{symbols}{"3F}

\begin{document}

\title{Randomized Aperture Imaging}

\author{Xiaopeng Peng}
\author{Garreth J. Ruane} 
\author{Grover A. Swartzlander, Jr.}\email{Corresponding author: gaspci@rit.edu}

\affiliation{Rochester Institute of Technology,54 Lomb Memorial Dr, Rochester, NY, 14623, USA}

\begin{abstract}
Speckled images of a binary broad band light source (600-670 nm), 
generated by randomized reflections or transmissions, were used to 
reconstruct a binary image by use of multi-frame blind deconvolution algorithms. 
Craft store glitter was used as reflective elements.
Another experiment used perforated foil.  Also reported here are
numerical models that afforded controlled tip-tilt and piston
aberrations.  These results suggest the potential importance of a 
poorly figured, randomly varying segmented imaging system.

\end{abstract}

\maketitle %% required

\section{Introduction} 
It is well known that angular resolution may be enhanced by use of a larger
aperture.  In practice, the size of a monolithic aperture is limited by the
cost, weight and construction constraints\cite{meinel1979cost,van2004scaling}.
Even if such constrains can be surmounted, adaptive optics methods must be
applied to achieve diffraction limited performance\cite{chanan2004control}.
Alternatively, passive approaches that make use of computer post-processing have
been successfully employed\cite{labeyrie1970attainment,fienup2013phase,chaudhuri2014blind}.  
Examples, including an aperture masking
system\cite{caroli1987coded}, and a multi-aperture
system\cite{duncan1999multi} have shown
great promise in astronomy and remote sensing\cite{chung2002design,miller2007optical,forot2007compton},biology\cite{haboub2014coded}, clinical trials\cite{bravin2013x,kavanagh2014feasibility}, and
new types of computational cameras\cite{veeraraghavan2007dappled,green2007multi,asifflatcam}. In
these imaging systems, the burden of hardware control are replaced or greatly
alleviated by digital computations. Mathematical tools, such as Fourier
analysis\cite{hariharan2003optical}, constrained optimization\cite{thiebaut2002optimization,bauschke2002phase}, 
and Bayesian inference\cite{ruiz2015variational} are essential in this
approach. Both aperture masking and multi-aperture systems have demonstrated an
improved signal-to-noise ratio of the acquired images, a calibrated point spread
function (PSF), rejection of atmospheric noise, and closure phase
measurements\cite{tuthill2006sparse}. However, the sparsity of apertures implies
a sparse coverage of spatial frequencies and loss of flux. Both adaptive optics
and aperture masking systems attempt to improve the images obtained from a
relatively well-figured optical system.  In this paper we address the case of an
ill-figured segmented optical system that varies randomly in time.  To make
matters worse, we assume no knowledge of the randomly varying PSF. We ask the
following the proof of concept question: Is it possible to reconstruct a near
diffraction-limited image from a series of recordings from such a system?

\begin{figure}[H]
\label{fig:fig_illus_system} 
\begin{center}
\includegraphics[width=1.0\linewidth]{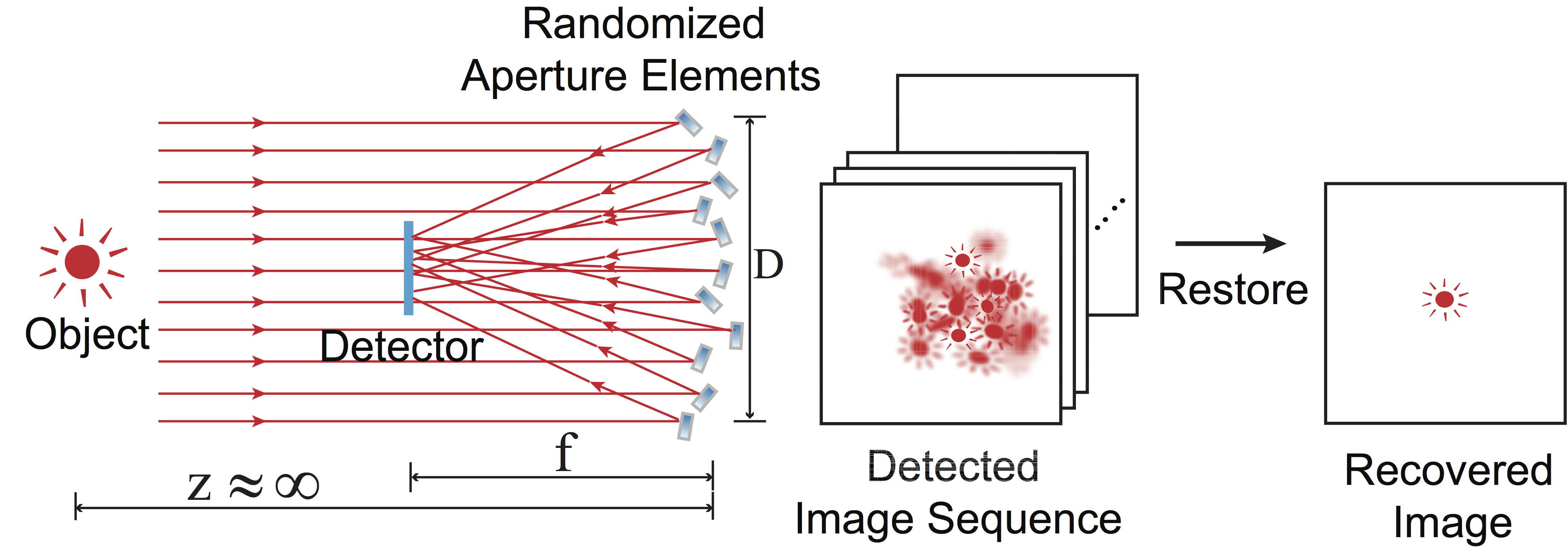} 
\end{center}
\caption{Random aperture imaging system having baseline $D$, depicting a distant 
light source, random elements, and a detector in an effective focal plane.
A sequence of images are recorded for different random orientations of the elements.
Computational imaging methods are used to reconstruct the image of the light source.}
\end{figure}

In some respects this study is related to the random lens imaging 
system\cite{fergus2006random}, where a collection of random reflectors served 
as a primary camera lens.  Similarly, the sparkle vision system\cite{zhang2014sparkle}, 
simplifies the random lens imaging system by using a lens to better focus the light. 
However, in these examples, the PSF was not time-varying, and intensive machine
learning algorithms were needed to uncover the PSF.  In contrast, we aimed to
reconstruct the time-varying PSF in a near real-time manner.

This report is organized as follows. In Section 2 we provide a description 
of a randomized complex aperture system.  Two different experimental scenarios
are established in Section 3: \lq\lq far field\rq\rq and \lq\lq near field\rq\rq.  For the purpose
of comparing experimental and modeled results, we next describe in Section 4
the corresponding numerical models.  Both the experimental and numerical data
are processed by means of the multiframe blind deconvolution scheme described
in Section 5.  We then report image construction results in Section 6.
Concluding remarks are provided in Section 7.

\begin{figure*}
\label{figure:fig_illus_exp_farfield} 
\begin{center}
\includegraphics[width=1\linewidth]
{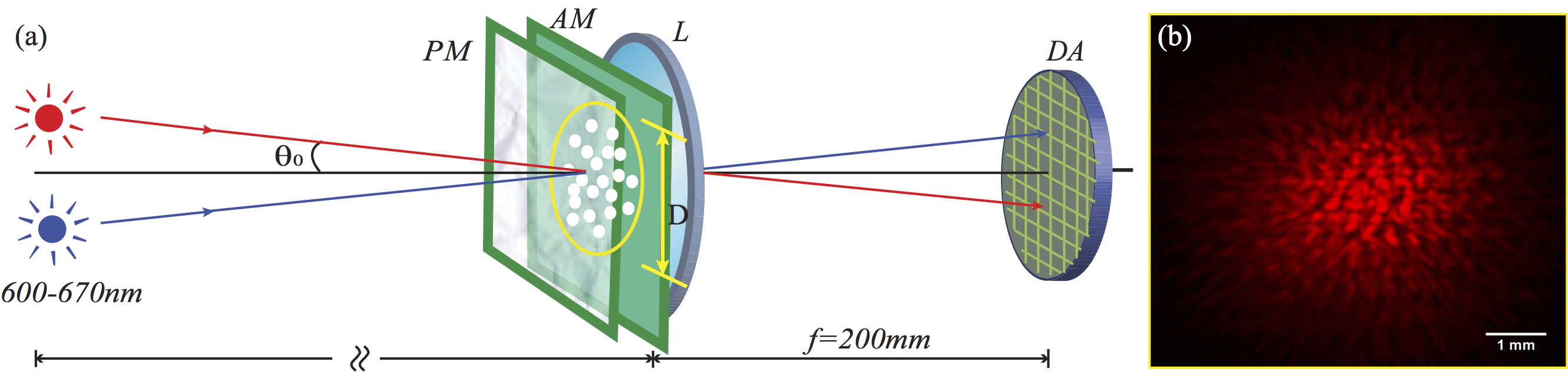} 
\end{center}
\caption{Configuration of the far field experiment. (a)Two polychromatic sources
subtend an angle $\theta_0$ with respect to the optical axis.  Rays are
transmitted though a phase mask ($PM$) and aperture mask ($AM$) near the front
face of a lens $L$ of focal length $f=200 mm$. Using different random masks,
multiple images are recorded on the detector array ($DA$), placed in the Fourier
transform (back focal plane) of the lens. (b) Example of a recorded image from a
random mask.}
\end{figure*}

\section{Basics of Randomized Aperture Imaging}  
An illustration of the randomized aperture scheme, shown in Fig. 1, depicts a
point source (or sources) an infinite distance from elements that are
distributed across a quasi-parabolic surface of baseline $D$ and a detector at
the mean focal distance $f$.  For a continuous parabolic reflector, a
diffraction-limited image would appear on the detector. However, tip,tilt and
piston errors associated with each reflecting element produces aberrated
speckles.  Further, the distribution may be evolving in time, providing a
sequence of $N$ speckled images. Here, we ignore motion blur. Postprocessing
techniques such a multi-frame blind deconvolution
\cite{jefferies1993restoration,harmeling2009online,hirsch2010efficient,kuwamura2014multiframe} 
may be used to recover a near-perfect image. This approach is particularly
suitable in cases where it is not practical to make repeated measurement of time
varying PSF.  What is more, multi-frame blind deconvolution is a self-heuristic
algorithm with less computational cost than machine learning, and may be simpler
to implement than phase retrieval methods when the PSF is time varying. An
advantage of this approach is (see Section 5) that whereas a single-image blind
deconvolution scheme prescribes a ratio of unknowns (the PSF and the recovered
object) to measurements as $2:1$, the multi-frame blind deconvolution scheme
improves the ratio to $N+1:N$. In general, the system shown in Fig. 1 may be
shift-variant, in which case shift-variant multi-frame blind deconvolution
algorithms\cite{hirsch2010efficient} are better suited. For the cases examined
below, however, the small shift-variance may be neglected, and we find the
shift-invariant multi-frame blind deconvolution algorithm is sufficient to
achieve good reconstructed image of a pair of point sources.

\section{Laboratory Demonstration}

\begin{figure*}
\label{fig:fig_illus_exp_nearfield} 
\begin{center}
\includegraphics[width=1\linewidth]
{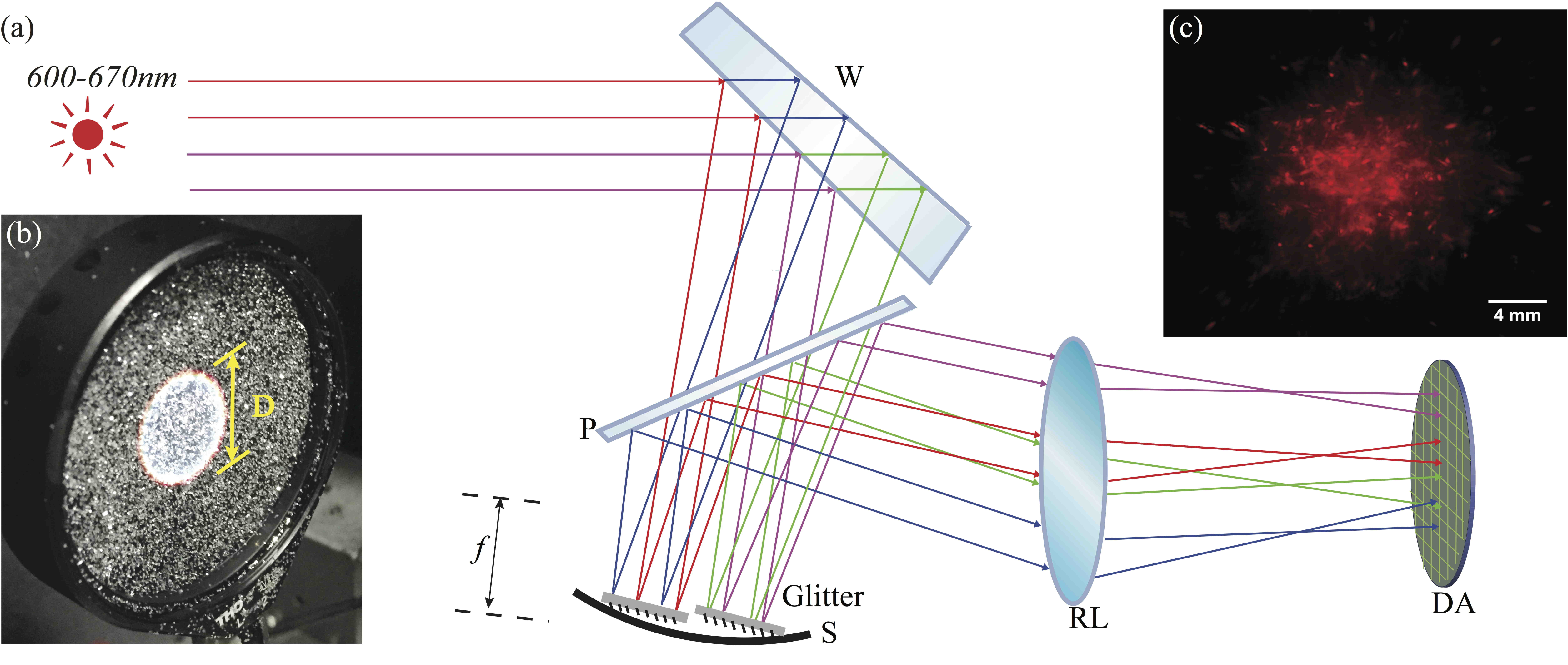}
\end{center} \caption
{Configuration of the near field experiment. (a) A glass wedge ($W$) is used to
create a double image of a single polychromatic light source.   The rays are
reflected from randomly distributed glitter across a blackened concave surface
($S$) having a radius of curvature $65.4mm$. A Pellicle beam splitter ($BS$)
direct the rays through a lens ($L$), which acts to relay the image at the focal
plane, to the detector array ($DA$). (b) Photograph of the random glitter
surface having an effective diameter, $D$. (c) Example of a recorded image of
the binary source from a random glitter surface. Multiple random images were
recorded by removing and then re-applying glitter.}
\end{figure*}

$Far$ $Field$. In the case shown in Fig 2, a \lq\lq far field" arrangement was
constructed whereby a transmissive
mask was positioned at the surface of a convex lens of focal length $f=200 mm$. A
set of $N = 50$ thin foil masks was produced, each having $M = 50$ randomly placed
pinholes(radius $r \approx 0.1 mm$ ) distributed across a $D = 3 mm$ diameter. The
foil was covered by a layer of wrinkled cellophane to randomize the phase at
each pinhole. The close proximity of the cellophane and foil allowed for a
nearly shift-invariant system. An Energetiq laser-driven white light source was
spatially and spectrally filtered to produce a collimated polychromatic beam
with a wavelength range $\lambda = 600-670 nm$. Light transmitted through the
mask was recorded at the back focal plane of the lens on the detector array of a
Canon 5D Mark III camera having a pixel pitch of $6.25 \mu m$ and detector size of $24
\times 36 mm$.  We call this arrangement \lq\lq far field" because the detected
light is governed by Fraunhofer diffraction from the pinholes.  Imaging
information is encoded in the interference of the diffracting beamlets. An
effective second mutually incoherent light source was introduced by transmitting
the beam through the system at angle $2 \theta_0$.  The two images were added
together to produce a single image of a binary source. The ground truth image
can be obtained using the same setup, but without the phase and aperture masks.

$Near$ $Field$. A schematic of the second experimental setup is shown in Fig 3.
for a \lq\lq near field" arrangement.  As above, spatially and spectrally
filtered light was formed into a collimated beam.  In this case, however,
reflecting elements were used to divert beamlets toward the detector.  We used
the front surface reflection from a glass wedge to produce a binary light
source. The reflecting elements were comprised of square \lq\lq fine size" craft store
glitter, with dimensions of roughly $0.3 \times 0.3 mm$.  The random aperture
condition was achieved by randomly sprinkling glitter across a blackened concave
surface having a radius of curvature $65.4mm$. After each image was recorded,
the glitter was washed off, and a new random surface was prepared.  The number
of sub-apertures across the full $D = 10 mm$ beam diameter ranged from $M = 200$ to 
400. The pellicle beam splitter was used to collect the reflected light and
direct it toward the detector array of the same camera.  Owing to space
constraints, the detector could not be placed directly in the focal plane of the
concave surface ($f=32.7 mm$ from the surface), and thus, a $150 mm$ relay lens
was used.  This configuration is called \lq\lq near field" because the beamlets from
each reflecting element undergo Fresnel diffraction upon reaching the detector. 
The diffraction length of a single reflecting element is roughly $450 mm$, which
is much greater than $f$. We
assert that the system is nearly shift invariant since the elements roughly
conform to a small patch of diameter $D$ on the concave surface. As seen in the inset of
Fig. 3, the elements display significant tip-tilt errors.

The reconstruction results of experimental data are discussed in
Section 6. Parameters of both schemes are listed in Table 1.

\begin{table} 
\caption{Values of Experimental Parameters} 
\label{table:table1_exp_nearfield}
\begin{center} 
\begin{tabular}{lll}
\hline Parameter & Far Field & Near Field\\ 
\hline 
       Number of sub-apertures in the cloud $M$   & 50  & 200-400\\
       Diameter of the baseline $D$[mm]& 3 & 10   \\ 
       Size of sub-aperture [mm]     & 0.2 & 0.3\\       
       Focal length $f$[mm] & 200 & 32.7\\
       Samples per $\lambda F \#$  in image plane[pix] & 7 & 0.4 \\
       Angular Separation of sources $\lambda /D $           & 26 & 625\\
       Bandwidth [$\Delta \lambda$]           & $\%10$ & $\%10$ \\ 
       Detector array size[mm$^{2}$] & 24 x 36\\
       Detector pixel pitch [$\mu$m] & 6.25 \\
\hline 
\end{tabular} 
\end{center} 
\end{table}

\begin{figure*}
\label{fig:fig_illus_num_farfield} 
\centerline{\includegraphics[width=0.92\linewidth]{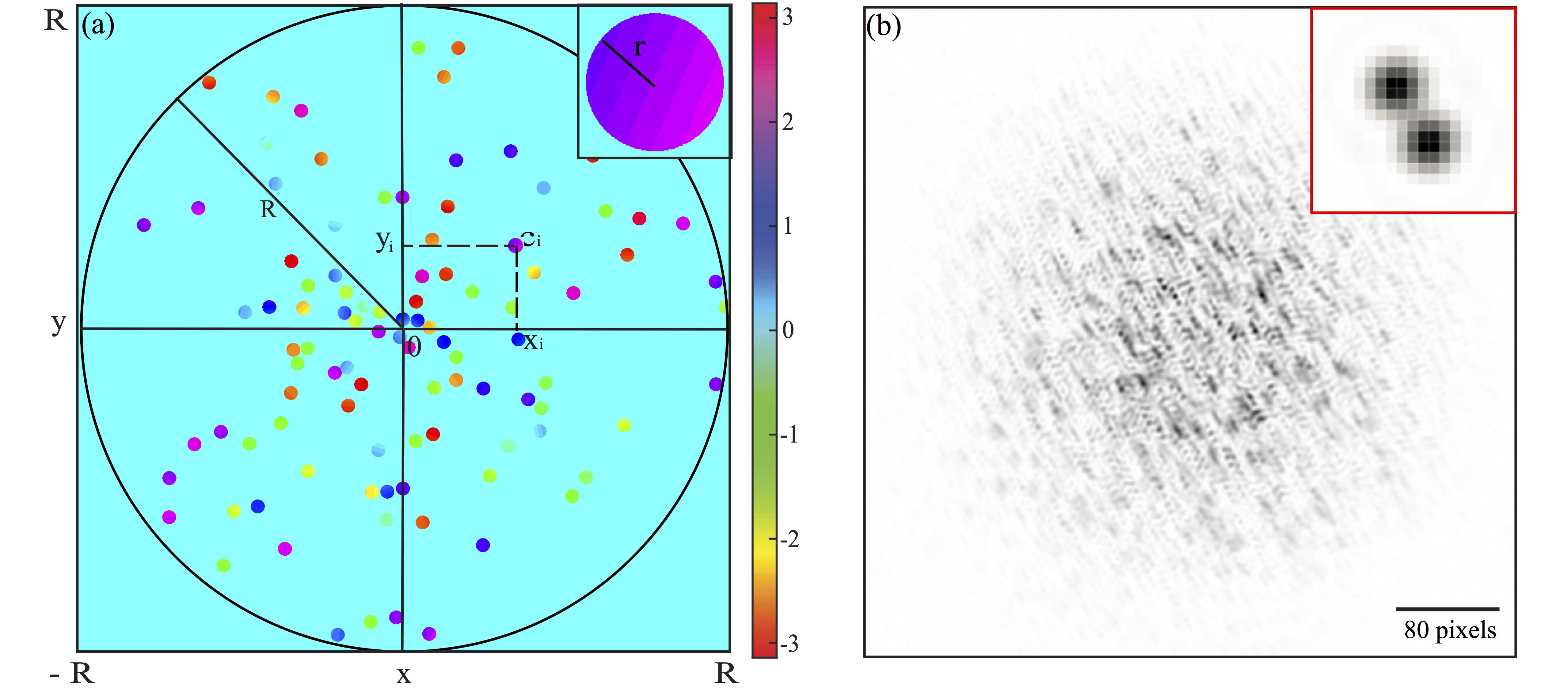}}
\caption
{Numerical model of the far field scheme.  
(a) Sub-apertures of radius $r$ (small circles) are distributed across the baseline 
radius $D=2R$ on an $N \times N$ grid ($N = 8192$). The transmitted light has a Gaussian 
random tip-tilt phase  (with standard deviation up to $40 \lambda / D$ across the 
diameter $D=2R$), 
as well as a uniformly distributed piston error of up to one wave.  The $M=100$ non-overlapping
randomly placed sub-apertures are assigned a position $\mathbf{c_{i}} = (x_i, y_i)$.
(b)Numerical image of a binary light source at the Fourier transform plane. 
The image has been inverted to aid the eye.} 
\label{fig:fig_NumericalFarField} 		
\end{figure*}

\section{Numerical Simulation of Optical System} 
 
$Far$ $Field$.  A numerical model of the far field experiment may be represented
by a distribution of complex circular sub-aperture functions 
(see Fig \ref{fig:fig_NumericalFarField}):

\begin{equation} 
U_{i}(\mathbf{x})=exp\left( \mathbf{| \tilde x_i |}\right)^{\beta}exp(i \phi_{i})
\end{equation}

\noindent where $\mathbf{\tilde x_i} = |\mathbf{x}-\mathbf c_{i}|_{2} / r_s$ is
a normalized vector,
$\mathbf{x} = (x,y)$ is an arbitrary vector in the plane of the aperture, 
$\mathbf{c_{i}} = (c_{x,i},c_{y,i})$ indicates the center of the 
$i$-th sub-aperture, $r_s$ is the radius of the sub-apertures, $|\cdot|_{2}$ is $l^{2}$-norm, and
$\beta =100$ is the power of Super-Gaussian function that defines the sub-apertures.
 
Phase aberrations, $\phi_i = \phi_{p,i} + \phi_{t,i}$, across each sub-aperture
are represented, respectively, by piston and tip-tilt terms:

\begin{equation} \phi_{p,i}(\mathbf{x}) = k a_{i}
\end{equation}

\begin{equation} \phi_{t,i}(\mathbf{x}) = k \mathbf{x}\mathbf{b}_i \end{equation}

\noindent where $k = 2 \pi / \lambda$ is wavenumber, $a_i$ is a pairwise independent uniformly distributed
random variable with interval equals to $\Delta z$, $\mathbf{b}_i$ is a pairwise independent Gaussian random 
variable with mean 0 and variance given by $<a^2_{i}> = \sigma^2_{0}$, and $<->$ denotes an ensemble average.

The complete system aperture function may be expressed as
\begin{equation}
U(\mathbf{x})=\left[\sum_{i=1}^{M}U_{i}(\mathbf{x})\right]exp\left(\dfrac{|\mathbf{x}-\mathbf{c}_0|_2}{R}\right)^\gamma
\end{equation}

\noindent where $M$ is the number of sub-apertures, $\mathbf{c}_{0} = (c_{x,0},c_{y,0})$ indicates the center of the 
baseline, and $R$  the radius of the full effective aperture, and $\gamma =100$ 
is the power of Super-Gaussian function that defines the baseline.

We further impose a non-overlapping condition to the sub-apertures:

\begin{equation} 
| \mathbf c_{i}-\mathbf c_{j} | > 2 r_s, i \neq j
\end{equation}

\noindent An example of the complex pupil function is shown in 
Fig \ref{fig:fig_NumericalFarField}.(a) for $M=100$ sub-apertures.

For two monochromatic point sources at infinity
having an apparent angular separation of $2\mathbf{\theta_{0}}$ along the x-direction, 
the electric field in the back focal plane of the
system may be expressed as Fourier transforms of the field from each source,
incident upon the imaging system(assuming it is shift-invariant):

\begin{equation} E^+(\mathbf{v}) = \frac{1}{\lambda} \sum_{\mathbf{x}}exp\left(i\frac{2\pi}{\lambda}\mathbf{v}\cdot\mathbf{x}\right)U(\mathbf{x})exp(+i k_x \mathbf{x})
\end{equation}

\begin{equation} E^-(\mathbf{v}) = \frac{1}{\lambda} \sum_{\mathbf{x}}exp\left(i\frac{2\pi}{\lambda}\mathbf{v}\cdot\mathbf{x}\right)U(\mathbf{x})exp(-i k_x \mathbf{x})
\end{equation}

\noindent where for paraxial rays, $k_x \approx 2 \pi \theta_0 / \lambda$. 

Assuming the two light
sources are mutually incoherent, we write the intensity in the plane of the
detector as: 

\begin{equation} I(\mathbf{x}) = |E^+|^{2}+|E^-|^{2} \end{equation}

\begin{figure*}[!hbt] 
\label{fig:fig_illus_num_nearfield} 
\begin{center}
\includegraphics[width=0.92\linewidth]
{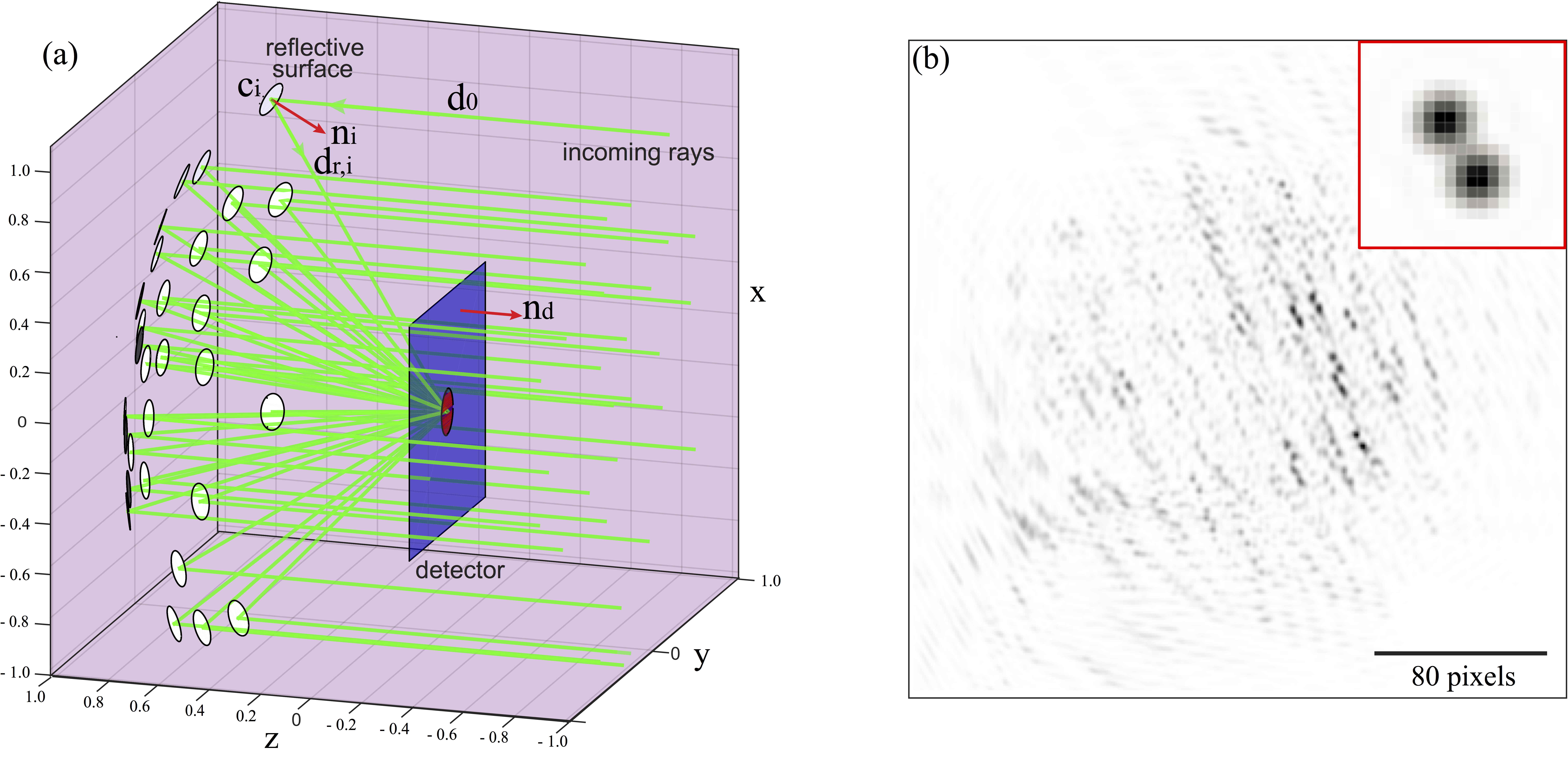} 
\end{center}
\caption
{Numerical model of the near field scheme.
(a) Randomly distributed circular reflecting elements of radius $r = 100$ pix
at the points $\mathbf{c}_i = (x_i, y_i, z_i)$ are aligned along a parabolic
surface with $F\#=75$, each provided with random tip-tilt
(with standard deviation up to $60 \lambda / D$ across 
the diameter $D =  20000$ pix), 
as well as a uniform piston error of up to ten waves.
Incoming collimated rays directed along the unit vector $\mathbf{d}_0$
are assumed to reflect in the direction $\mathbf{d}_{r,i}$
without diffracting as they propagate to the focal plane detector array.
Respective element and detector surface normals: $\mathbf{n}_i$  and $\mathbf{n}_d$.
(b) Example of a numerical image of a binary light source at the focal plane.} 
\end{figure*} 

$Near$ $Field$ In the near field regime, the numerical
model ignores diffraction of the reflected light as it propagates
from the aperture to the focal plane detector.  That is, we assume the characteristic
diffraction distance is much longer than the effective focal
length: $\pi r_s^2 / \lambda >> f$.  

The geometric model of near field scheme is illustrated in Fig 5.
For a single source of collimated light, we assume a beam of parallel ray
aligned along the unit vector $\mathbf{d}_{0}$. Rays incident upon the $i_{th}$
reflector centered at the point $\mathbf{c}_{i}$ are reflected along the unit
vector $\mathbf{d}_{r,i}$:

\begin{equation} 
\mathbf{d}_{r,i} = \mathbf{d}_{0} - 2\left(\mathbf{d}_{0}\cdot\mathbf{n}_{i}\right)\mathbf{n}_{i}
\end{equation} 

\noindent where the Cartesian components of the unit normal vector of the $i_{th}$ reflector 
may be expressed in terms of direction cosines as:
$\mathbf{n}_{i} =(\cos \mathbf{\theta}_{i})$, 
where $\mathbf{\theta}_{i} = (\theta_{x,i}, \theta_{y,i}, \theta_{z,i})$. 

If the reflectors conformed to a paraboloid 
\begin{equation}
z = -(x^2+y^2)/4f+f
\end{equation} 
\noindent where $f$ is
the focal length. Each of the $i_{th}$ rays directed along $\mathbf{d}_{r,i}$
would coincide at the focal point $\mathbf{o}=(0,0,0)$.  In that special case
the direction angles of the normal vector of each reflector may be expressed as  
$\mathbf{\theta}_{i,0} = (\theta_{x,i,0}, \theta_{y,i,0}, \theta_{z,i,0})$, where

\begin{equation}
    \begin{split}
       \theta_{x,i,0} = cos^{-1}(x/2pf)\hspace{-4.8mm}
	\\ \theta_{y,i,0} = cos^{-1}(y/2pf)\hspace{-4.2mm}
	\\ \theta_{z,i,0} = cos^{-1}(1/p) 
	\end{split}
\end{equation}
 
\noindent and $p = ((x/2f)^2+(y/2f)^2 + 1)^{1/2}$. We note theses angles are related by 
$ cos^2(\theta_{x,i,0}) + cos^2(\theta_{y,i,0}) + cos^2(\theta_{z,i,0}) =1$.

In general, the normal vector of each reflector will suffer from random tip-tilt errors 
$\mathbf{\Delta \theta_{i}} = (\theta_{x,i}\Delta a_{x,i} a_{y,i}\Delta \theta_{y,i},  a_{z,i}\Delta\theta_{z,i})$, 
and thus the directional angles becomes
\begin{equation}
\mathbf{\theta}_{i} = \mathbf{\theta}_{i,0} + \mathbf{\Delta \theta_{i}}
\end{equation}
\noindent where $a_{x,i}, a_{y,i}, a_{z,i}$ are pairwise independent Gaussian random variables
variable with mean 0 and variance equal to $\sigma_0$.

Piston error, $w_{i}\Delta z_{i}$ may be introduced by displacing the center of the $i_{th}$
mirror from the paraboloid surface defined by 
$\mathbf{c}_{i}= (c_{i,x}, c_{i,y}, f + (c_{i,x}^2 + c_{i,y}^2)/4f)$, where $w_{i}$ is pairwise independent 
uniform random variable with interval of $\Delta z_{i}$

Let us consider a ray incident upon an arbitrary point $\mathbf{x} = (x,y,z)$ on
the $i_{th}$ flat reflecting element. The plane of the element is represented by the
equation $\mathbf{n}_{i}(\mathbf{x} - \mathbf{c}_{i})=0$. For a circular reflector 
of radius $r_s$ we impose $|\mathbf{x} -\mathbf{c}_{i}|^{2}\le {r_{s}}^{2}$. 
The intersection points of the incident beam with the plane of the $i_{th}$ circular reflector 
can be represented parametrically as:

\begin{equation} 
\mathbf{m}_{i}= \mathbf{c}_{i} + \alpha\cos t \mathbf{e}_{1} + \alpha\cos t\mathbf{e}_{2}
\end{equation} 

\noindent where $t\in [0, 2\pi]$, $\alpha \in[0, r_{s}]$, 
$\mathbf{e}_{1}$ and $\mathbf{e}_{2}$ are unit vectors that satisfy
$\mathbf{e}_{1} = (0,1, cos \theta_{x} / cos \theta_{y})/(1+(cos \theta_{x} / cos \theta_{y})^2)^{1/2}$ 
and $\mathbf{e}_{2}=\mathbf{e}_{1} \times \mathbf{n}$.

Next we wish to determine where the reflected beam intersects the detector plane.
In general, the detector plane can be defined by a center point $\mathbf{q}$ (usually 
the focal point of the paraboloid) and a normal vector $\mathbf{n}_{d}$. The projection of the points $\mathbf{m}$
along the reflected beam onto the detector plane may then be expressed as:
 
\begin{equation} \mathbf{u}_{i} =\mathbf{m}_{i}+
\mathbf{d}_{r} \frac{(\mathbf{q}-\mathbf{m}_{i})\cdot \mathbf{n}_{d}} 
{\mathbf{d}_{r}\cdot \mathbf{n}_{d}} 
\end{equation} 

The electric field within each of the $i_{th}$ projected beams in the detector
plane may be represented as a tilted plane wave scaled by an obliquity factor
from both the mirror $\mathbf{d}_{0} \cdot \mathbf{n}_{i}$, and the detector
$\mathbf{d}_{r} \cdot \mathbf{n}_{d}$.

With two point sources at infinity, we repeat the above procedure by
including two bundles of incoming rays reflected from an element along 
unit vectors $\mathbf{d}^+_{0}$ and $\mathbf{d}^-_{0}$ respectively. 
The two fields in the detection plane are labelled $E^+_{i}$ and $E^-_{i}$:
\begin{equation}
    \begin{split} 
E^+(\mathbf{u}^+_{i}) = A^+_i exp(i k \mathbf{d}^+_{r,i} \cdot \mathbf{u}^+_{i})exp(i k L^+_{i})\\\
E^-(\mathbf{u}^-_{i}) = A^-_i exp(i k \mathbf{d}^-_{r,i} \cdot \mathbf{u}^-_{i})exp(i k L^-_{i})
\end{split}
\end{equation}

\noindent where $A^+_i$ and $A^-_i$ are zero valued outside the projection area of the $i_{th}$
mirror. $L^+_i$ and $L^-_i$ are full path length of rays that travel from the binary source
to the $i_{th}$ mirror, and then reflected onto the detector. $k=2\pi/\lambda$ is the wavenumber. For small angular deviations in the detector plane and for 
equally luminous point sources, we make the approximation of setting the 
$A^+_i$ and $A^-_i$ within the interior regions of each bundle of rays equal for all $\mathbf{u}_{i}$.  Here we consider two mutually incoherent 
sources subtending an angle $2 \theta_0$ 
and bisecting the z-axis:  $\mathbf{d}^+ \cdot \mathbf{d}^- = cos(2 \theta_0)$.
Defining the postion vector on detector plane as $\mathbf{x}_{d} = (x_d,y_d)$, the measured irradiance may be expressed:

\begin{equation} 
I(\mathbf{x}_{d}) = |\sum_{i=1}^{M} E{_i}^+(\mathbf{x}_{d})|^2 +|\sum_{i=1}^{M}E{_i}^-(\mathbf{x}_{d})| ^2
\end{equation}

The ground truth image is captured by imaging the objects using a monolithic
mirror of the baseline size. The value numerical parameters are given in Table 2. 

\begin{table} 
\caption{Value of Numerical Parameters} 
\label{table:table1_num_farfield}
\begin{center} 
\begin{tabular}{lll}
\hline Parameter & Far Field & Near Field \\ 
\hline 
Image plane grid size [pix] & 4096  & 4096  \\
Radius of sub-aperture $2r$[pix]& 82 & 100\\
Diameter of the baseline $D$[pix]& 8092 & 20000\\
Pixels per $\lambda F\#$ in image plane & 4 & 7.5\\
Angular separation of the objects[$\lambda/D$] & 1.5 & 1.5 \\ 
\hline 
\end{tabular} 
\end{center} 
\end{table} 

\section{System Estimation and Image Recovery}
Multi-frame blind deconvolution algorithms are used to recover a target 
scene from a set of blurry, noisy and distorted observations. 
They are generally categorized into two types:
(1) non-blind deconvolution, where the target scene is reconstructed based on
complete or partial knowledge of the point spread function(PSF) of the imaging
system; and (2) blind deconvolution, where the unknown target scene and system point spread
function are recovered simultaneously without a priori knowledge.
Since the work of Ayers and Dainty\cite{ayers1988iterative}, multi-frame blind deconvolution has become 
an important tool for  image recovery, resulting in
numerous research efforts and applications.
Common approaches involve: (1) Batch mode multi-frame
blind deconvolution\cite{jefferies1993restoration,peng2015image}, where all the distorted
observations are processed at the same time; and (2) Serial mode
multi-frame blind deconvolution\cite{harmeling2009online,hirsch2011online,peng2014mirror}, where degraded
inputs are sequentially processed. 
Compared with the batch mode, the serial approach is
more memory efficient and can in principle be done at the same time as the image
acquisition.  In this paper image restoration was accomplished by use of 
a serial multi-frame blind deconvolution\cite{hirsch2011online} scheme.

Here we give a brief review of the online multi-frame blind deconvolution algorithm. 
We assume that at each time point
$n=1,2,\cdots,N$, the random aperture mirror system records a blurred image
$\{g_{n}\}$. Assuming the imaging system is shift-invariant(or approximate
shift-invariant), the image formation process can be modeled as convolution of
the target image and the system PSF, where the recorded $n_{th}$ image $g_n (x,y)$ may
be expressed:

\begin{equation} g_{n} = (f \otimes h_{n})+q_{n} \end{equation} 

\noindent where $\otimes$ denotes the two-dimensional convolution operator, and
$f(x,y)$ and $q_n(x,y)$ respectively represent the ideal
image and its random noise, and $h_n(x,y)$ represents the shift-invariant PSF of the
$n_{th}$ time frame.  We assume the PSF changes from frame to frame owing to time
varying tip, tilt, and piston errors, as well as the location of each mirror.
The goal for multi-frame blind deconvolution is to recover the ideal image $f$
and the temporally varying PSF $h_{n}$ from a set of degraded images
$\{g_{n}\}$. A simple but effective choice for the blind deconvolution can be
achieved by solving the following non-negatively constrained problem \cite{harmeling2009online}

\begin{equation}\label{eq:eq_costfun} \{f,h_{n}\} = \min_{h_{n}\geq 0,f\geq
0}\frac{1}{N^{2}}\sum_{n=0}^{N} \sum_{\mathbf
u}|{G_{n}}-\tilde{F}\tilde{H_{n}}|^{2} (\mathbf u) \end{equation}

\noindent where $\tilde{G_{i}}, \tilde{F}, \tilde{H_{i}}$ are the Fourier
transforms of the observed images $g_{i}$, ideal image $f$, and PSFs $h_{i}$
respectively, and $\mathbf u$ is the position vector in frequency space.

Solutions to Eq.\ref{eq:eq_costfun} are commonly solved using either: 
(1) batch
mode optimization using the constrained conjugate gradient descent of cost
function with respect to $f$ and $\{h_{n}\}$ in an alternating
manner\cite{jefferies1993restoration,kuwamura2014multiframe,peng2015image}; and
(2) serial mode optimization using the multiplicative
updates\cite{harmeling2009online,hirsch2011online}, as is shown in the method we
employed, outlined below in Algorithm 1,estimated image and PSF are updated
respectively using multiplicative method in each iteration:

\begin{equation} H_{i} = H_{i-1}\odot
\frac{F_{i}^{T}G_{i}}{F_{i}^{T}(F_{i}H_{i-1})} \end{equation}

\begin{equation} F_{i} =  F_{i-1}\odot
\frac{H_{i}^{T}G_{i}}{H_{i}^{T}(H_{i}F_{i-1})} \end{equation} \noindent 
and where $\odot$ denote the component-wise multiplication. 

\removelatexerror 
\begin{algorithm}[H] 
\SetAlgoLined 
\caption{Alternating Minimization for Online Blind Deconvolution} 
\KwIn{Captured image sequence
$\{G_{n}\}$} 
\KwOut{Restored image $F$ and PSFs $\{H_{n}\}$} 
Initialize $F =F_{0}$, $H_{0}$\\ 
$i=1$\\ 
\While{$\sum_{\mathbf u}|{G_{i}}-{F}{H_{i}}|^{2}>\epsilon$ and $i<N$ } 
{ 
\hspace{7mm}$H_{i}=\arg\min\sum_{\mathbf u}|{G_{i}}-{F}_{(i-1)}{H}|^{2} $ \;
\hspace{7mm}$F_{i}=\arg\min\sum_{\mathbf u}|{G_{i}}-{F}{H_{i}}|^{2}$\;
\hspace{7mm}$i = i+1$ \;
\hspace{6.5mm}$F = F_{i}$ } 
\end{algorithm}

\section{Results and Discussion} 
Analyses of our reconstructed images resulting from multi-frame blind deconvolution
are presented below.  First we describe the experimentally measured data for far field
and near field schemes using a polychromatic (10$\%$ bandwidth) light source.  
Numerically generated data is then used to demonstrate the
reconstruction of a monochromatic binary light source under different 
tip-tilt and piston errors.  Quantitative 
comparisons between ground truth and reconstructed images are evaluated based
on two metrics: spatial separation error $E_{D} = |D-D'|/D$ 
and the peak intensity ratio error 
$E_{p} = |I_{p,1}/I_{p,2} - I'_{p,1}/I'_{p,2}| / (I_{p,1}/I_{p,2})$,
where $D$ and $D'$ are the respective distances between the peaks of 
the ground truth and restored images respectively, and
where $I_p$ and $I'_p$ are the respective peak intensities of the ground truth and 
restored results.

\subsection{Experimental Results}
Reconstructed images were obtained by use of $N=50$ images for 
both the far field and near field schemes. 
Details of the experimental setup and parameters are discussed in Section 4 and
Table 1. The restored images of a binary light source and examples of
speckle images are shown in Fig 6. Immediately we see in both cases
that the reconstructed images are superior to the speckle images, i.e, qualitative agreement
with the ground truth is acheived.  Quantitatively, we find
the distance between the intensity peaks are in good agreement, with $E_{D} \approx 5\%$.
Furthermore the intensity peaks are equal to within $E_{p} \approx 15\%$.
This is remarkable considering the $10\%$ bandwidth of the light source, and the
estimated $15 \lambda /D$ tip-tilt error and a likely piston error of at least several waves.
The good agreement between the ground truth and reconstructed images may be attributed
to the high degree of shift invariance of the imaging systems.  That is, the speckle 
data in both Fig 6.(b) and (d) contains multiple overlapping pairs of binary images.  We note
that in the far field (diffracted) case the pairs displaced from the central region of the
speckle image are diffused owing to the broad bandwidth of the light source.  
In contrast, for the near field case, the beamlets from each reflecting element
experience little diffraction, and thus no chromatic spreading of the beamlets.  
We believe the multiframe BD scheme is successful
at recovering the binary light source in both cases because the underlying imaging systems
are well described by a shift invariant convolution process.

\begin{figure*} 
\label{fig:fig_exp_farfield} 
\begin{center} 
\includegraphics[width=1\linewidth]
{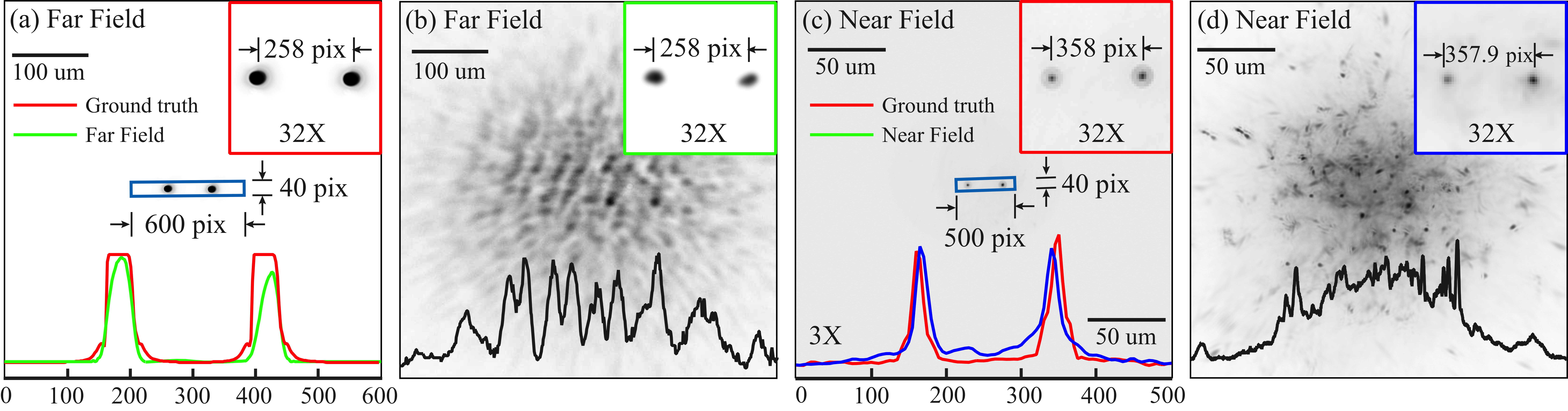} 
\end{center} 
\caption{Image reconstruction results of multi-frame blind deconvolution applied to an
experimental far field sequence and an experimental near field sequence
respectively(polychromatic with band width $\Delta \lambda/\lambda= 10\%$). The far field sequence is obtained with $M = 50$ non-overlapping 
sub-apertures, and the near field sequence is obtained with $M \approx 200-400$ non-overlapping 
reflecting elements. The tip-tilt error in near field sequence is approximate to $15\lambda/D$ 
(a)Groundtruth image of the binary light sources in far field scheme, and the comparison of the 
line profile of the ground truth and the recovered result; (b)A typical image in the
captured far field sequence and the enlarged part of the recovered result; (c)Groundtruth image of the binary light sources in near field scheme, and the comparison of the 
line profile of the ground truth and the recovered results; (d)A typical image in the
captured near field sequence and the enlarged part of the recovered result; }
\end{figure*}

\subsection{Numerical Results}
To explore how well multiframe blind deconvolution restores a binary image from random
aperture mirror images, numerically simulated far field and near field data were
generated for various degrees of tip-tilt and piston error.
Details of the system were described in Sec 4 and Table 2.

First we examine numerical cases that closely resemble the experimental system.
For both the near and far field schemes we numerically generated speckled images
with 50 non-overlapping sub-apertures.  Gaussian random tip-tilt error with $\sigma_{0}=10 \lambda/D$
and uniform random piston error with $\Delta z = 1 \lambda$ were assumed for each of 
the $N=60$ images of the generated sequence.  Shown in Fig 7 are the ground truth and
example speckle images for the far field and near field scenarios.  The insets show
a zoomed image of the ground truth in Fig 7.(a), and the corresponding reconstructed
images in Fig 7.(b,c).  As in the experimental case, the qualitative agreement between
the ground truth and reconstructed images is good. Quantitatively the spatial
separation error is $E_{D} \approx 10\%$ for both cases, and the magnitude ratio
error is $E_p$ is less than $4\%$.  These errors are smaller than the experimental
values because the numerical cases are noise free and monochromatic.  We found that
the near field values of $E_D$ and $E_p$ are somewhat larger than the far field values.
We attributed this difference to the lower degree of shift invariance in the near field 
case owing to different projections of the beamlets on the detector plane. 

To examine the fidelity of reconstructed image as a function of tip-tilt and
piston error, we modeled the far field system with $\sigma_{0}=10,20,40[\lambda/D]$
with no piston error (see Fig 8), and then $\Delta z = 0.5,1.0,1.5\lambda$ 
with no tip tilt error (see Fig 9).  In these
cases we set the number of sub-apertures to $M=100$, and the number of
images to $N=50$.
As expected, the quality of the reconstructed images deteriorates with increasing
phase error.
The reconstruction errors increase from $E_{D} \approx
8.9$, $9.1$ to $9.4\%$, and $E_{P} \approx 2$, $24$ to $34\%$
as the tip-tilt error increase from $\sigma_{0}=10$ to $20$ to $40[\lambda/D]$. 
On the other hand, as the increment of piston error increases from $\Delta z = 0.5$ to
$1.0$ to $1.5\lambda$, the errors for the restoration increase from 
$E_{D} \approx 13$, $16$ to $22\%$, and $E_{P}$ increase from 
$\approx 6$, $27$ to $58\%$. 
For both tip-tilt and piston error we find the magnitude ratio error suffers more significantly
than the spatial separation error. We attribute this to the increased intensity of speckles and 
interference patterns that are resulted from sufficiently large phase errors.

\begin{figure*}
\label{fig:fig_num_farVSnear} 
\begin{center}
\includegraphics[width=0.75\linewidth]
{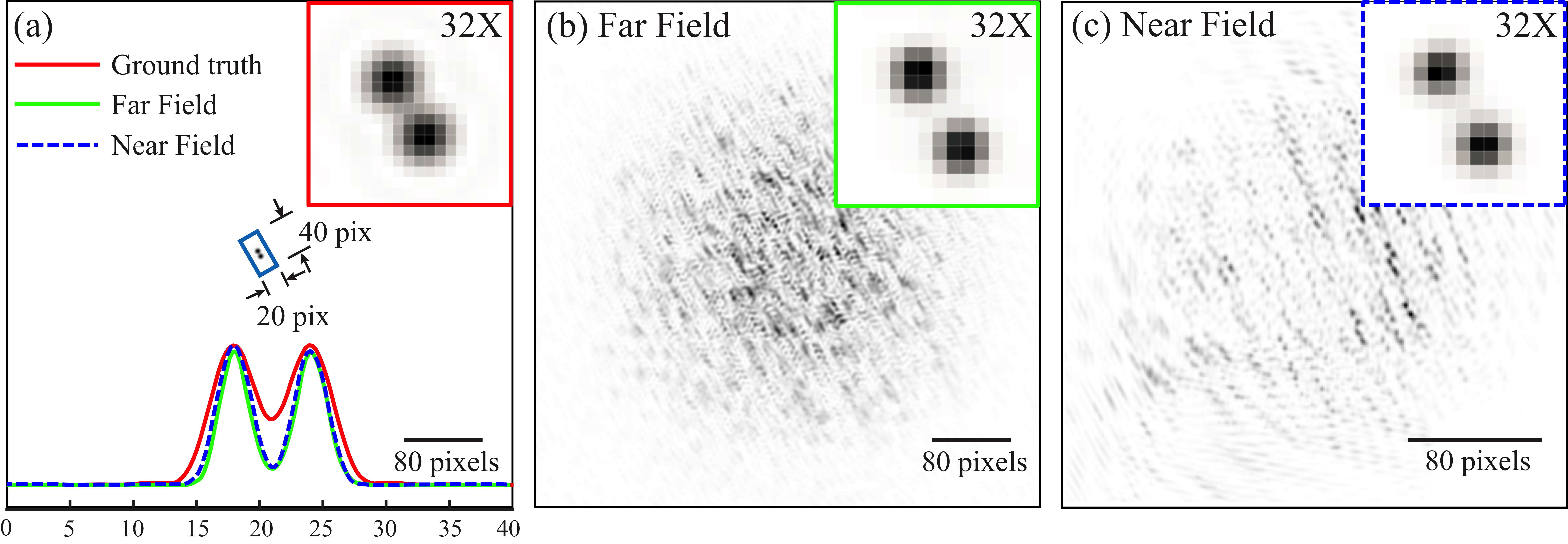} 
\end{center} 
\caption
{Comparison of image reconstruction results of multi-frame blind deconvolution applied to a
numerically simulated far field sequence and a near field sequence
respectively(monochromatic). Both sequences are generated with $M = 50$
non-overlapping sub-apertures(or reflecting elements), and having random tip-tilt error of $\sigma_{0}=10
\lambda/D$ and piston error of $\Delta z = 1 \lambda$. Each sequence consists
of $N \approx 60$ images. (a)Groundtruth image of the binary light sources, and the comparison of the 
line profile of the ground truth and the recovered results; (b)A
typical far field image in the sequence, and enlarged part of recover result;
(c))A typical near field image in the sequence, and enlarged part of recover
result}
\end{figure*}

\begin{figure*}
\label{fig:fig_num_farfield_tiptilt}  
\begin{center}
\includegraphics[width=1\linewidth]
{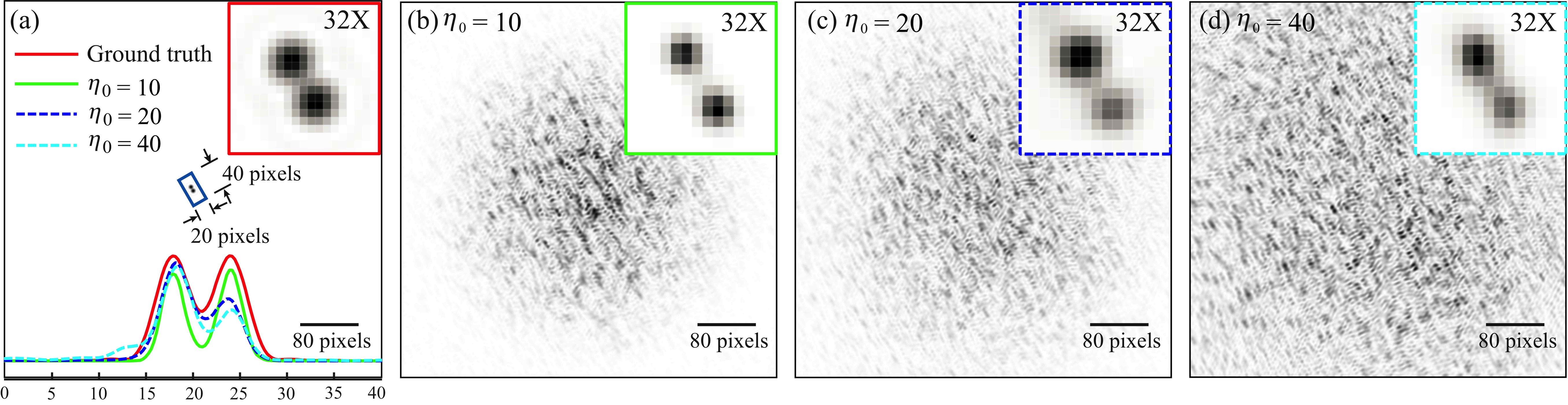} 
\end{center} 
\caption{Comparison of tip-tilt error to the image reconstruction results. Three monochromatic far
field sequences are numerically simulated with $M=100$ non-overlapping sub-apertures having random 
tilt-tilt error with $\sigma_{0}$ of 20, 40, and 60$[\lambda/D]$ respectively, and 
piston phase error with of $\Delta z = 1 \lambda$.Each sequence consists of $N = 60$ images.(a)Groundtruth image of the binary light sources;
(b)-(d) a typical image from each of the three sequences and enlarged part of its reconstruction result. } 
\end{figure*}

\begin{figure*}
	\label{fig:fig_num_farfield_piston}  
	\begin{center}
		\includegraphics[width=1\linewidth]
		{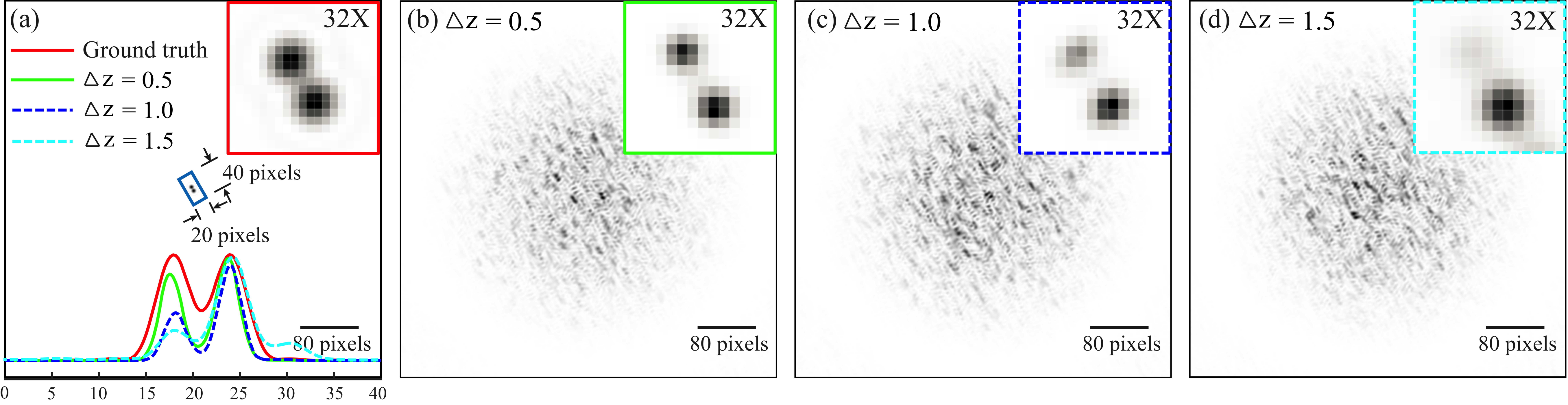} 
	\end{center} 
	\caption{Comparison of piston error to the image reconstruction results. Three monochromatic far
		field sequences are numerically simulated with $M=100$ non-overlapping sub-apertures having random 
		tilt-tilt error with $\sigma_{0} = 40\lambda/D$, and piston phase error with $\Delta z = 0.5, 1.0, 1.5[\lambda]$ respectively.
		(a)Groundtruth image of the binary light sources, and comparison of line profile of the groundtruth and reconstructed results;
		(b)-(d)a typical image from each of the three sequences and enlarged part of its reconstruction result. } 
\end{figure*}

\section{Conclusion} 
We numerically and experimentally explored the concept of random aperture mirror
telescope for both monochromatic binary light sources and polychromatic binary
light sources, in both far field and near field schemes. For an approximate
shift-invariant system, binary light sources can be restored using Multi-frame
blind deconvolution techniques from both experimental and numerical data with
reconstruction error kept in a tight tolerance. The numerical results further
demonstrate that for an approximate shift-invariant system, a near diffraction
limit resolution ($1.5\lambda F\#$) can be achieved in the presence of tip-tilt
of 40$\lambda/D$ and piston phase up to 1.0$\lambda$ for monochromatic sources.

Several interesting aspects remain yet to be analyzed. We would like to
quantifying the phase errors by the use of spatial light modulator in the
experiments. Also, further investigation of a few system
parameters, i.e. fill factor of sub-apertures, varying $F\#$, the number of
light sources, the magnitude ratio among sources, as well as noise will be
conducted for both experiments and numerical simulation to better evaluate
the system performance.

\section{Acknowledgment} This work was funded by the US National Science
Foundation (ECCS-1309517) and by the NASA Innovative Advanced Concepts (NIAC)
program.  We are grateful to Alexandra B. Artusio-Glimpse (RIT) for valuable
suggestions about the experiment.  We are also very grateful to Dr. Marco B. Quadrelli
(NIAC Principal Investigator) and Dr. Scott A. Basinger (Jet Propulsion Laboratory) 
for discussions about future space telescopes.

\end{document}